\documentclass[final,3p]{elsarticle}

\usepackage{graphicx}

\usepackage{amssymb}
\usepackage{lineno}
\usepackage{lscape}
\usepackage{tabularx}
\usepackage[gen]{eurosym}
\usepackage{multirow}
\usepackage{hhline}
\usepackage{booktabs}
\usepackage{longtable}
\usepackage{siunitx}
\usepackage{framed}
\usepackage{todonotes}
\usepackage{wasysym}

\journal{Information and Software Technology}

\begin{document}

\begin{frontmatter}

\title{Non-Technical Individual Skills are Weakly Connected to the Maturity of Agile Practices}

\author[chalmersgu]{Lucas Gren\corref{cor1}}
\ead{lucas.gren@cse.gu.se}
\cortext[cor1]{Corresponding author. Tel.: +46 739 882 010}

\author[chalmersgu]{Alessia Knauss}
\ead{alessia.knauss@chalmers.se}

\author[holland,holland2]{Christoph Johann Stettina}
\ead{c.j.stettina@cdh.leidenuniv.nl}

\address[chalmersgu]{Chalmers University of Technology and the University of Gothenburg, SE-412 96 Gothenburg, Sweden}

\address[holland]{Centre for Innovation, Leiden University,
Schouwburgstraat 2, 2511 VA, The Hague, The Netherlands 
}

\address[holland2]{LIACS, Leiden University,
Niels Bohrweg 1, 2333 CA, Leiden, The Netherlands}

\begin{abstract}

\textit{Context}
Existing knowledge in agile software development suggests that individual competency (e.g.\ skills) is a critical success factor for agile projects. While assuming that technical skills are important for every kind of software development project, many researchers suggest that non-technical individual skills are especially important in agile software development.

\textit{Objective}
In this paper, we investigate whether non-technical individual skills can predict the use of agile practices.

\textit{Method}
Through creating a set of multiple linear regression models using a total of 113 participants from agile teams in six software development organizations from The Netherlands and Brazil, we analyzed the predictive power of non-technical individual skills in relation to agile practices. 

\textit{Results} 
The results show that there is surprisingly low power in using non-technical individual skills to predict (i.e.\ explain variance in) the mature use of agile practices in software development. 

\textit{Conclusions}
Therefore, we conclude that looking at non-technical individual skills is not the optimal level of analysis when trying to understand, and explain, the mature use of agile practices in the software development context. We argue that it is more important to focus on the non-technical skills as a team-level capacity instead of assuring that all individuals possess such skills when understanding the use of the agile practices.

\end{abstract}


\end{frontmatter}

\section{Introduction}
Agile methods are increasingly used in industry as they are established to support projects in their success \citep{serrador2015does}. \citet{cockburn2001agilepeoplefactor} argue that individual competency is an important success factor in agile projects. In agile methods ``the emphasis on people and their talent, skill, and knowledge becomes evident.'' Even on team-level, they argue that the emphasis is again ``on competency rather than process.'' Literature suggests that we progress through two major stages during the development of a cognitive skill, a declarative knowledge stage and a procedural knowledge stage \citep{anderson1982acquisition}. While the former can be acquired by reading text books (e.g.\ learn how to lead a team), the latter, the procedural knowledge, can only be acquired in process (e.g.\ by actually leading a team and learning from mistakes).
Hence, it seems that success in agile projects depends on individual skills, that are developed in individuals over time. Many studies in software engineering have focused on explaining the individual skills (see e.g.\ \citet{turley1994identifying}), which implies that the individual non-technical skills are believed to predict team-level performance in relation to collaborative aspects. However, no studies have looked at explicit connections between agile practices and individual non-technical skills. As we assume technical skills to always be a precondition for a successful software development endeavour, non-technical skills seem to be important in such endeavours as well \citep{Bender2014}. Such non-technical individual skills have been stated as especially important in agile practices, since they focus more on individuals than processes \citep{cockburn2001agilepeoplefactor}. 

Therefore, this paper investigates the connections between (i.e.\ the predictive power of) thirteen self-assessed non-technical individual skills and the intended mature use of a set of eight common agile practices. In this paper, we define \emph{practice maturity} by assuming it to be possible to describe by an evolutionist model comprising of a progressive and directional set of changes. As the practices mature, they increase in perfection or complexity over time \citep{king1984evolution}. We assume the end goal of the development of the agile practices to be to what degree a practice is implemented in its intended way, as measured by the degree of agreement to items in the \emph{perceptive agile measurement} \citep{so}. We did not investigate the use of agile practices by having subjects tick the practices they use from a list, but instead used the perceptive agile measurement \citep{so} to assess the degree of agile behavior prescribed by an agile practice. We assume different degrees of such agile behavior to be equivalent to different levels of agile practice maturity. The perceptive agile measurement was created to assess the social-psychological effect of agile practices, which provides a higher resolution of the actual behaviour shown in these agile teams.

The sample consisted of agile team members from seven organizations in Brazil and The Netherlands. We set up eight hypotheses regarding the associations between all individual skills and each agile practice. In order to assess the hypotheses, we conducted a diversity of analyses: First, we checked our survey data for normality. Second, we ran eight independent ANOVAs. Third, we built three new regression models for the significant models and analyzed their effect sizes. 

The results show non-significant or negligible effect sizes for all our analyses. We therefore conclude that looking at non-technical individual skills is not the optimal level of analysis when wanting to understand the use of agile practices, and argue for using the ``agile team'' as the level of analysis instead.

To clarify the outline of this paper, the next section (Section~\ref{sec:related_work}) presents related work with regards to individual and team level skills in agile software development. Section~\ref{constructs} gives an overview of the measurements and constructs (i.e., non-technical individual skills and agile practice) used in this paper. Section~\ref{sec:methodology} depicts the method used to measure agility and non-technical individual skills and correlate these measurements in a set of multiple linear regression models, Section~\ref{sec:results} presents the results, which are discussed in Section~\ref{sec:discussion}, followed by conclusions and future work in Section~\ref{sec:conclusions}.

\section{Related Work}\label{sec:related_work}  
In this section, we provide a definition of what skills are, provide an overview of research on individual skills in the software engineering context, and present the few studies looking at a higher level of analysis in relation to skills in the software engineering domain. 

\subsection{What Is a Skill?}\label{skillsdef}
We define a skill as ``an ability to do something well; expertise'' \citep{dictionary} in this study in its broader sense. However, in order to understand what this means, we need to look at related research in what characterizes a skill and how they are acquired.
As mentioned in the introduction, the acquisition of a cognitive skill can be described by two major stages: a declarative knowledge stage and a procedural knowledge stage \citep{anderson1982acquisition}. These stages are intimately connected to what \citet{argyris2000double} calls single- and double-loop learning. We can learn to repeat new information without integrating it on a deeper level. With such shallow learning, we can not apply our acquired knowledge in new situations and we fail to translate the new knowledge to similar cases \citep{argyris2000double}. Therefore, we asked the participants of this study about their own satisfaction in relation to specific skills, which then are more in relation to the procedural knowledge (double-loop) rather than shallower declarative (single-loop) knowledge in their work context. In this study, we also define a team-level skill as the capability of the entire team in relation to non-technical skills, i.e.\ even if all members can not plan well for themselves, the team as an entity might have good planning skills anyways as a product of collaboration and reciprocal help.

Another scientific field that has gone through similar phases in research is the medical field, and \citet{fletcher2002role} shows how surgical teams have gone from only focusing on technical skills to also realizing the importance of the non-technical skills needed for successful treatment of patients. \citet{fletcher2002role} divides the non-technical skills into two categories; (1) cognitive and mental skills (e.g.\ decision-making, planning, etc.) and (2) social or interpersonal skills (e.g.\ team-working, communication, leadership, etc.). This research highlights the importance of individual non-technical skills, which has also been common in the software engineering research and is presented next.

\subsection{Research on Individual Skills in Agile Software Development}\label{skills}
Agile methods are described as having a strong focus on individuals and their skills \citep{cockburn2001agilepeoplefactor}. \citet{cockburn2001agilepeoplefactor} conclude that individual skills seem more important than team characteristics -- even on the team level they argue that the emphasis is on individual competency.
Strengths as well as weaknesses of individual skills need to be taken into consideration as both can have an influence on the success of an organization. \citet{conboy2011peopleoverprocess} report an increased exposure of capabilities and reliance on social skills. They report that exposing weaknesses of team members can often be counter-productive and even highly respected and performing team members can be bullied, challenging existing organizations. Furthermore, an understanding of the skills necessary within a team can help team members in their development.

In agile software development, the individual skills of software engineers need to be considered, according to many studies. For example, \citet{turley1994identifying} identified 38 essential competencies of software engineers. Among the top ten are: (1) Team Oriented, (2) Seeks Help, (3) Helps Others, (4) Use of Prototypes, (5) Writes\slash Automates Tests with Code, (6) Knowledge, (7) Obtains Necessary Training\slash Learning, (8) Leverages\slash Reuses Code, (9) Communication/Uses Structured Techniques for Communication, and (10) Methodical Problem Solving. 
%
%
%
In addition to the general skills that a software engineer should possess, skills of requirements engineers play an important role in agile software development. The focus in agile software development is on the customer who decides on what is of value and who is supposed to be on site to clarify requirements \citep{Fricker2010}. Hence, contrary to traditional RE, in agile development any member from the development team can directly interact with the customer and collect requirements \citep{Sillitti2005}. 
Furthermore, RE represents an area in which it is especially important to consider social\slash non-technical skills and theories \citep{Malley2004}. Hence, non-technical skills of requirements engineers are important and required for each individual team member in agile software development.
%

Negotiation is an inevitable element in RE \citep{Nuseibeh2000, Grunbacher2005}. For example in requirements elicitation and analysis it plays a major role as it supports handling
stakeholder conflicts concerning requirements \citep{Ahmad2011}. Negotiation promotes a shared vision, shared knowledge, and cooperation among stakeholders \citep{Ahmad2011}. 
Furthermore, communication is an inevitable skill in RE \citep{Coughlan2003}. 
 Team members need to communicate with each other for many reasons, i.e. bugs discussion, code issues, code reviews, code refactoring, code synchronization, coordination,
management, support issues, sprint planning, quality, user story
clarification and user story negotiation \citep{Inayat2015a}.
In agile requirements engineering most activities depend on communication between different parties, their input and judgment. Hence, it is highly dependent on the skills of the team members. Further skills for phased as well as agile RE are: 1) Dividing bigger tasks into small ones, 2) giving up control -- as code is developed and changes by different people, 3) writing meaningful tests, 4) conversation, 5) object-oriented design, 6) fast cycle times \citep{Kovitz2003}.
\subsection{Research on Team-Level Skills in Agile Software Development}\label{org}
Most existing research focuses on individual skills, as presented in Section~\ref{skills}. A few studies also report about the importance of non-technical skills on a team level (e.g., with regards to team work and setting up teams).  \citet{Lalsing2012} identified the underlying people factors for a team to be effective in agile software development, and several agile projects were in the study. Results showed that for projects exceeding the project budget, issues were e.g.\ related to team communication and collaboration (i.e.\ trust and interaction). They concluded that it is crucial to select the ``right people for the right team,'' and not only the ``right people.''
\citet{Tanner2014} study success and failure of agile projects in waterfall environments. From studying two cases of agile projects and literature on this topic, they concluded that the following factors have an influence on the projects success or failure of agile projects: 1) Culture, 2) Customer Involvement and Mandate, 3) Stakeholder Involvement and Buy-In, 4) Team Structure and Team Logistics, 5) Project Type and Project Planning, and 6) Skill Level and Attitude of Team Members. 
\citet{Crowder2015} identified communication, coordination, trust and team orientation as the most important team factors for distributed agile teams, based on survey data. From a systematic literature review, they identified team orientation, shared leadership, mutual performance monitoring, backup behavior, feedback, team autonomy, team learning, coordination, communication, trust, collective culture, ease of use of technology, and team familiarity as teamwork factors. \citet{grenjss2} also found that team maturity (from a group dynamics perspective) is a key factor in the success of building agile teams in large organizations. 

A challenge when trying to understand productivity in general is to get data from the right level of analysis \citep{hackman2003learning}. Software engineering research on human factors can learn a great deal from the journey social psychology has done from the individual to the group as a level of analysis (cf.\ \citet{hogg2000we}), a journey that has largely been repeated by organizational science \citep{klein2000micro}. If we want to understand agile practices, the individual skills level might be too much on a micro level based on research in other fields, and we could study the agile teams and how software developers and other team-members behave collectively instead, i.e.\ explained variance in data might come from the meso level of the team as an entity. Recent research in social psychology \citep{woolley2010evidence,engel2014reading} has shown that, in general, the performance of teams on a diversity of tasks is set on group-level independent of the intelligence of the individuals. The intelligence of groups have instead been shown to be more dependent of social sensitivity (i.e.\ a person's ability to read emotions in facial expression), and conversational turn-taking (i.e.\ groups were less collectively intelligent if a few individuals dominated the conversations). Such findings contradict the conclusions drawn in a lot of software engineering research on the importance of non-technical individual skills (see for example \citet{turley1994identifying}). To clarify such contradictions, we conduct a study in the software engineering domain aiming to shed light on this contradiction of individual vs.\ group level. As software engineering mainly looks at individual level, this is the level we chose to investigate in this current study. According to \citet{hackman2003learning}, one should preferably look at the macro (i.e.\ organizational) level in addition to the micro and meso levels. While we recognize that variance possibly could be explained on the organizational level, we focus on discussing the micro and meso levels in this study.

\section{Measurements, Constructs, and Research Hypotheses}\label{constructs}
In this section we describe the measurements used in the current study and their operationalization based on previous studies. We first present theory on common agile practices and how they can be measured, and then we suggest a measurement of self-assessed non-technical individual skills based on such studies in the agile context.

\subsection{Common Agile Practices}
Agility as a concept can be difficult to delineate \citep{laanti2013definitions}. Agility emerged in practice and has been discussed across different scientific domains such as manufacturing and logistics \citep{boothharmer1994agile}, business management \citep{vanOosterhout2010business}, information systems literature \citep{Conboy2009}, and sports science \citep{sheppard2006agility}, which makes it difficult to define. Fuelled by the difficulties in definition, an analysis by \citet{laanti2013definitions} suggests to look at agility as a set of concrete practices to understand agility. Other studies have followed that track and investigated the usage and perceptions of practices perceived as agile within software development teams \citep{williams,so}. In line with \citet{salvato2009capabilities} we believe that such concrete, routinized activities have a huge impact on the effectiveness and sustainability of project management processes -- and, as such, that there must be some kind of behaviour that can be considered ``more'' agile than other behaviour connected to more traditional project management groups. Early Scrum literature generally describes to the following practices \citep{schwaberBeedle2003agile}: (1) Collocated scrum teams, 
(2) Daily scrum stand-up meetings
(3) Iteration planning in sprint planning meetings, 
(4) Iterative development in sprints, and 
(5) Sprint reviews. 

Based on previous research and perceptions of practitioners, \citet{so} constructed an instrument for quantitative analyses of social-psychological effects of the above mentioned agile practices, however categorized a bit differently than those Scrum practices above. They created scales for the eight of the core agile practices, namely (1) Iteration planning, (2) Iterative development, (3) Continuous integration and testing, (4) Stand-up meetings, (5) Customer access, (6) Customer acceptance tests, (7) Retrospectives, and (8) Collocation. The framework of \citet{so} is particularly useful as they provide for the first time a scientifically validated psychometric instrument covering these eight core agile practices, i.e.\ their tool tries to capture agile behaviour in connection to the practices and therefore claims to measure the actual practices. The measured practices Customer access, and Customer acceptance tests, also extend their survey to include aspects of agile requirements engineering, which adds a focus on the entire development chain from planning to delivering in the agile context. 

We use the questionnaire suggested by \citet{so} to understand the connection of agile practices and individual skills with the reason that we need social-psychological measurements in our study.
In the following, we will elaborate on the eight practices included in the instrument by \citet{so}. Furthermore, in the text boxes below each of the depicted practice we reproduce the exact items (i.e., the agile practice and sub-questions used to cover this agile practice) used in our questionnaire.

\paragraph{Iteration Planning} In a Collaborative planning workshop the deliverables and scope of an iteration is defined with  all team members being present. It is executed at the beginning of each iteration, sometimes also referred to as \emph{Planning Game}. The practice generally consists of two stages: In the first stage, requirements are gathered in the form of user stories to serve as a medium for discussions between the customer and the developers. In the second stage, the stories are revised, estimated and prioritized into an iteration backlog \citep{liu2005environment}. The active participation of technical team members in definition as well as estimation of user stories is considered as an indication for a mature application of the practice \citep{wang2012assimilation,sharp2004ethnographic}.

\begin{framed}
\scriptsize \textbf{Iteration Planning:} (1) All members of the technical team actively participated during iteration planning meetings. (2) All technical team members took part in defining the effort estimates for requirements of the current iteration. (3) When effort estimates differed, the technical team members discussed their underlying assumption. (4) All concerns from team members about reaching the iteration goals were considered. (5) The effort estimates for the iteration scope items were modified only by the technical team members. (6) Each developer signed up for tasks on a completely voluntary basis. (7) The customer picked the priority of the requirements in the iteration plan.
\end{framed}

\paragraph{Iterative Development} Routinized delivery of sub-results (working software) in short and iterations of fixed length~\citep{petersen2}. Although the practice has been popularized with the dawn of agile methods, the application of iterative software development dates as far back as the mid-1950s \citep{larman2003iterative}. Short iterations of 30 days or less together with continuous integration, have been found as the two practices considered most essential for a team to be considered agile \citep{williams}. Iterative development is a shared practice in agile methods as well as in user-centered design~\citep{chamberlain2006towards}.

\begin{framed} 
\scriptsize \textbf{Iterative Development:} (1) We implemented our code in short iterations. (2) The team rather reduced the scope than delayed the deadline. (3) When the scope could not be implemented due to constraints, the team held active discussions on re-prioritization with the customer on what to finish within the iteration. (4) We kept the iteration deadlines. (5) At the end of an iteration, we delivered a potentially shippable product. (6) The software delivered at iteration end always met quality requirements of production code. (7) Working software was the primary measure for project progress.
\end{framed}

\paragraph{Continuous Integration and Testing} \citet{holck2003continuous} define continuous integration as follows: (1) access of development team members to add contributions to the development version at any time, and (2) obligation of team members to integrate their own contributions properly. In order to enable such a continuous integration of ongoing development into a software system, the practices are ofter linked to (automated) testing methods to enable a timely verification of the system \citep{hellmann2012agile}.

\begin{framed}
\scriptsize \textbf{Continuous Integration and Testing:} (1) The team integrated continuously. (2) Developers had the most recent version of code available. (3) Code was checked in quickly to avoid code synchronization/integration hassles... (4) The implemented code was written to pass the test case. (5) New code was written with unit tests covering its main functionality. (6) Automated unit tests sufficiently covered all critical parts of the production code. (7) For detecting bugs, test reports from automated unit tests were systematically used to capture the bugs. (8) All unit tests were run and passed when a task was finished and before checking in and integrating. (9) There were enough unit tests and automated system tests to allow developers to safely change any code.
\end{framed}

\paragraph{Stand-up meetings} Frequent team coordination meetings in which team members provide a status update to their colleagues. The meetings are generally hold standing up and are time boxed to 5-15 minutes to frame its short and focused nature. Each coaching session starts with a team stand-up where each group was asked the three common questions: ``What have you done since the last meeting?'' ``What are you planning on doing until the next meeting?'' and ``What issues and impediments are you facing that prevent you from accomplishing these things?'' An ethnographic account of the practice is provided by \citet{sharp2004ethnographic}. \citet{stray2012investigating} investigated the application of daily stand-up team coordination meetings. They found that only 24\% of each of the meetings they studied focused on coordination. Rather, 35\% of the meeting time was spent on content-discussions elaborating problem issues and discussing possible solutions.

\begin{framed}
\scriptsize \textbf{Stand-Up Meetings:} (1) Stand up meetings were extremely short (max. 15 minutes). (2) Stand up meetings were to the point, focusing only on what had been done and needed to be done on that day. (3) All relevant technical issues or organizational impediments came up in the stand up meetings. (4) Stand up meetings provided the quickest way to notify other team members about problems. (5) When people reported problems in the stand up meetings, team members offered to help instantly.
\end{framed}

\paragraph{Customer Access} Availability of customers for product feedback and clarification of requirements is integral to effective agile teams and has been found as one of the critical success factors when implementing agile methods \citep{misra2009identifying,chow2008survey}. Especially when moving away from traditional software development customer access can be a challenge as customers might not be used to close interaction \citep{misra2009identifying}.

\begin{framed}
\scriptsize \textbf{Customer Access:} (1) The customer was reachable. (2) The developers could contact the customer directly or through a customer contact person without any bureaucratic hurdles. (3) The developers had responses from the customer in a timely manner. (4) The feedback from the customer was clear and clarified his requirements or open issues to the developers.
\end{framed}

\paragraph{Customer Acceptance Tests} Acceptance tests defined by the customer present a means to the developers to determine which iteration goals have been achieved at the end of each iteration \citep{so2010making}.

\begin{framed}
\scriptsize \textbf{Customer Acceptance Tests:} (1) How often did you apply customer acceptance tests? (2) A requirement was not regarded as finished until its acceptance tests (with the customer) had passed. (3) Customer acceptance tests were used as the ultimate way to verify system functionality and customer requirements. (4) The customer provided a comprehensive set of test criteria for customer acceptance. (5) The customer focused primarily on customer acceptance tests to determine what had been accomplished at the end of an iteration.
\end{framed}

\paragraph{Retrospectives} Workshop at the end of each iteration to improve the process and incorporate successful practices for the next iteration. In order to enable continuous improvement of the practices applied each team member lists ``What went well'' and ``What could be improved'' \citep{maham2008planning}. The impact, however, dependent largely on their implementation \citep{mchugh2012agile}. For example, retrospectives should take place at the end of each iteration and systematically assign all improvement points to responsible individuals \citep{so2010making}. 

\begin{framed}
\scriptsize \textbf{Retrospectives:} (1) How often did you apply retrospectives? (2) All team members actively participated in gathering lessons learned in the retrospectives. (3) The retrospectives helped us become aware of what we did well in the past iteration(s). (4) The retrospectives helped us become aware of what we should improve in the upcoming iteration(s). (5) In the retrospectives (or shortly afterward), we systematically assigned all important points for improvement to responsible individuals. (6) Our team followed up intensively on the progress of each improvement point elaborated in a retrospective. 
\end{framed}

\paragraph{Collocation} Close proximity of development team members is reported as one of the critical success factors when implementing agile methods \citep{lindvall2002empirical}. It is recognized as one of the important vehicles for successful communication and knowledge creation \citep{misra2009identifying}.

\begin{framed}
\scriptsize \textbf{Collocation:} (1) Developers were located majorly in... (2) All members of the technical team (including QA engineers, db admins) were located in... (3) Requirements engineers were located with developers in... (4) The project/release manager worked with the developers in... (5) The customer was located with the developers in...
\end{framed}

\subsection{Common Non-Technical Skills}
We use common denominators on skills derived from previous studies as presented in Section \ref{skills}. In that analysis, we found 13 non-technical individual skills that have been identified in at least two papers as important for software developers to be successful. Table \ref{skillstable} presents the non-technical skills used in this study and supporting literature depicting the importance of this skill.

\begin{table}
\renewcommand{\arraystretch}{1.5}
\caption{Non-technical individual skills used in the present study}
\label{skillstable}
\centering
\begin{tabular}{ll}
\hline
\bfseries  Non-technical skills & \bfseries Studies including skill    \\
\hline
(1) Communication Skills & \citep{lee2006managers, turley1994identifying, Coughlan2003, Lalsing2012, Kelle2015, napier2009projectskills, Crowder2015}  \\
\hline
(2) Teamwork Skills &  \citep{lee2006managers, turley1994identifying, cockburn2001agilepeoplefactor, Tanner2014, napier2009projectskills, Crowder2015}   \\
\hline
(3) Collaboration Skills & \citep{turley1994identifying, Lalsing2012, Crowder2015}  \\
\hline
(4) Ability to Meet Project Goals &  \citep{turley1994identifying, chen2001validation, Crowder2015}    \\
\hline
(5) Customer Orientation & \citep{turley1994identifying, Fricker2010, Kelle2015, napier2009projectskills, Tanner2014}   \\
\hline
(6) Requirements Management Skills & \citep{Sillitti2005, Ahmad2011}  \\
\hline
(7) Planning Skills & \citep{lee2006managers, turley1994identifying, napier2009projectskills, Tanner2014}     \\
\hline
(8) Leadership Skills & \citep{lee2006managers, Kelle2015, napier2009projectskills, Crowder2015, Cohn2004} \\
\hline
(9) Decision-Making Skills & \citep{napier2009projectskills, janis1977dma}   \\
\hline
(10) Business-Minded Skills & \citep{turley1994identifying, Crowder2015}   \\
\hline
(11) Problem-Solving Skills & \citep{turley1994identifying, napier2009projectskills}   \\
\hline
(12) Organizing Skills & \citep{lee2006managers, turley1994identifying}   \\
\hline
(13) Negotiation Skills & \citep{turley1994identifying, Grunbacher2005, Ahmad2011, Shell2001, Crowder2015}   \\
\hline

\end{tabular}
\end{table}

\subsection{Research Hypotheses}

Based on the literature, in this current paper, we investigate the assumption that the use of agile practices is positively connected to non-technical individual skills. All the items were self-assessed by team members in agile software development projects. The hypotheses were therefore the following:

\begin{itemize}
\item \textbf{H$_1$:} The mature use of the agile practice Iteration Planning is positively associated with agile team members' self-assessed non-technical skills. 
\item \textbf{H$_2$:} The mature use of the agile practice Iterative Development is positively associated with agile team members' self-assessed non-technical skills. 
\item \textbf{H$_3$:} The mature use of the agile practice Continuous Integration and Testing is positively associated with agile team members' self-assessed non-technical skills. 
\item \textbf{H$_4$:} The mature use of the agile practice Stand-Up Meetings is positively associated with agile team members' self-assessed non-technical skills.
\item \textbf{H$_5$:} The mature use of the agile practice Customer Access is positively associated with agile team members' self-assessed non-technical skills.
\item \textbf{H$_6$:} The mature use of the agile practice Customer Acceptance Tests is positively associated with agile team members' self-assessed non-technical skills.
\item \textbf{H$_7$:} The mature use of the agile practice Retrospectives is positively associated with agile team members' self-assessed non-technical skills.
\item \textbf{H$_8$:} The mature use of the agile practice Collocation is positively associated with agile team members' self-assessed non-technical skills.
\end{itemize}

\section{Method}\label{sec:methodology}
In order to investigate the connections between the individual skills and agile practices, we created a survey using the agile practices suggested by \citet{so} and our aggregation of non-technical individual skills (both derived in the previous section).

\subsection{Participants}
All the participants were members of agile teams in the participating companies (as stated by our company contacts). We explicitly asked all team members to answer the survey, but to skip questions they were not able to assess. As a consequence, most respondents were intimately connected to development code in one way or the other. The reason why we did not asked specifically for only software developers, were that some employees that conduct software development might have another title, like for example ``system engineer.''

This study was carried out at six organizations in total, three companies in Brazil and three in The Netherlands (two companies and one public sector IT department). These companies were selected because the first and third authors had direct or indirect research connections to people within the organizations. We wanted the participating organizations to be as diverse as possible in order to being able to generalize our study to the broader population of agile team members in the software development context, i.e.\ we intended to survey agile team members from different continents, different sizes of organization, as well as the public and private sectors (see Table \ref{participants}). 

\begin{landscape}
\begin{table}
\renewcommand{\arraystretch}{1.5}
\caption{The Participating Organizations}
\label{participants}
\centering
\begin{tabular}{cccccccccccc}
\hline
\bfseries  Organization & \bfseries Country &\bfseries  \# of employees  &\bfseries  \# of responses  \\
\hline
Company 1 & Brazil & ~100   & 22  \\
\hline
Company 2 & Brazil & ~5,000   & 57   \\
\hline
Company 3 & Brazil & ~35   & 11  \\
\hline
Company 4 & The Netherlands & ~3,500   & 4  \\
\hline
Company 5 & The Netherlands & ~50,000   & 7  \\
\hline
Public Sector Organization & The Netherlands & ~5,000  & 12    \\
\hline

 &  &   &   Total: 113* \\
\end{tabular}
\end{table}
*After list-wise deletion due to missing values, we had at least 99 valid responses to use in our regression analyses.

\end{landscape}

The Brazilian sub-sample contained data points from IT departments at a large on-line media and
social networking enterprise with around 5,000 employees, a smaller software consultancy company with around 35 employees, and a company that provides programming courses to individuals and companies, with around 100 employees. All the agile team members received the surveys via their managers but their replies were anonymous. The response rate was 92\% for the Brazilian sub-sample due to the fact that they were collected on-site in paper form.

The Dutch sub-sample consisted of four groups across three organizations: an IT service provider (around 3,500 employees), banking and financial services (around 50,000 employees), and an IT service department in the public sector (around 5,000 employees). The response rate for the Dutch sub-sample was 81\%.

All the participating companies said they use an agile development approach in the software development conducted, but with teams of different maturity levels in that process. The total number of respondents in the first measurement was 158 agile team members.


\subsection{Survey items}

In the survey these skills were separately put as questions in the following form: ``How satisfied are you with your [skill]?'' The reason why we used personal satisfaction of a certain non-technical skill was that, in agile teams, such skills are perceived as utterly important. Therefore, a satisfaction of a skill should be more related to a perceived peer-evaluation by a team-member than if asked to only rate their own individual skills. In addition, as mentioned previously, we aimed at investigating procedural rather than declarative skills, and in order for a participant to rate their individual skill high in context, we believe such rating will be more in relation to their declarative knowledge. 

We used the survey suggested by \citet{so} to measure the mature use of the eight agile practices included in their study, i.e.\ higher scores on their survey imply higher maturity of a practice since the questions are in relation to the intended, and therefore mature, use of a practice. The entire construct used is presented in Section~\ref{constructs}. Due to all the different definitions and ambiguity of ``agility'' \citep{laanti2013definitions}, we chose this survey since it captures the social-psychological behavior in connection to what the different practices try to achieve. It is also the only tool we have found that is validated through a factor analysis \citep{fabrigar} and a reliability analysis (using the Cronbach's $\alpha$ \citep{cronbach}) with a sample of $N=227$ \citep{so}.

The agile items in the questionnaire were assessed on a 7-point Likert scale (1 = never and 7 = always), with one exception being the Collocation items that were rated from 1 = the same room to 5 = different timezones. These scales were used for the simple reason that these measurements were developed and validated using those exact scales. The skills were assessed from 1 = completely dissatisfied and 9 = completely satisfied. 

In order to validate our results further, we also built a regression model using perceived code quality as a response variable. The perceived code quality aspect was measured using the single question: ``How would you rate the code quality in your product(s)?'' rated from 1 (very poor) to 4 (excellent). The reason for this validation was to investigate if non-technical individual skills were connected to a completely different aspect of the software developed (i.e.\ other than the agile practices). If that were the case, we would have support for the usefulness of measuring non-technical individual skills outside the scope of agile practices.

\subsection{Data collection and analysis}
The questionnaires were distributed in paper form by one of the managers and collected on site by the first and third authors (depending on the country). 

\begin{figure}
\centerline{\includegraphics[scale=0.5]{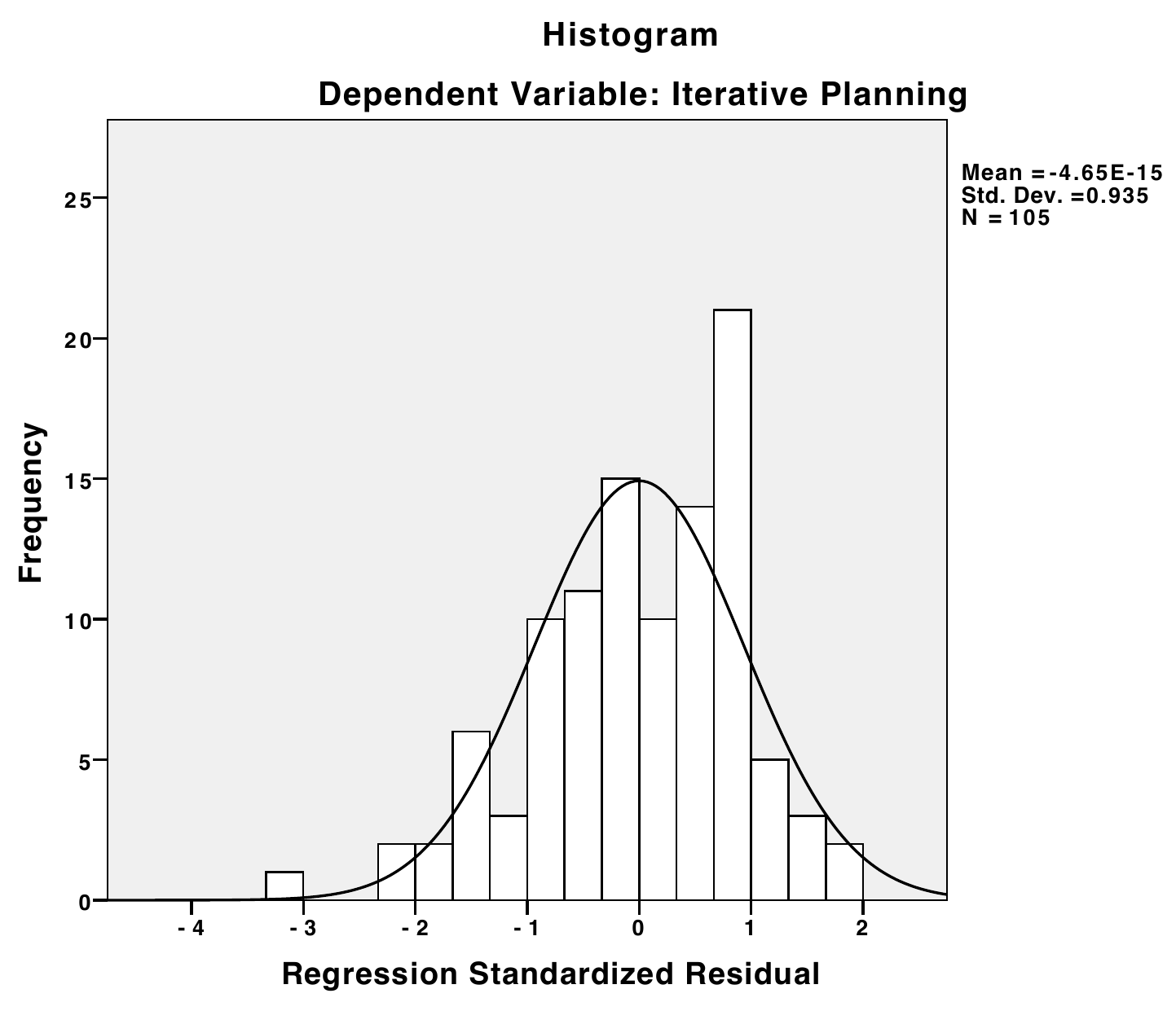}}
\caption{Frequency histogram with Iteration Planning as dependent variable and all the individual skills as factors.}
\label{d}
\end{figure}

To evaluate if the data was normally distributed, we plotted frequency histograms for all the multiple linear regression models (for one example, see Figure~\ref{d}). Figure~\ref{d} shows that the residuals are randomly scattered around the regression line. However, we saw some issues with the dependent variable Retrospectives, which turned out to have a set of outliers. When these outliers were removed and the data looked normally distributed, the ANOVA for that category was still not significant, meaning that the outliers did not affect the result. We also checked the variance inflation factor (VIF) for each regression model and all values were below 3, which is acceptable for these kinds of analysis since a common rule of thumb is below 10 \citep{marquaridt1970generalized}.

In order to investigate the connections between the two concepts we first ran eight ANOVAs with all the skills as factors and the agile practices as response variables one by one. The purpose was to investigate the predictive power of knowing the agile team members perceived individual skills on all the agile practices separately (plus the quality question in its own analysis as a validation). This means that we investigated how much of the variance in a measured agile practice that was explained by the non-technical individual skills together, and therefore, opted to use linear regression analysis with all the skills as factors. It is important to note the differences between predictive and causal models and in this study we only claim the former. However, since we have the theoretical assumption that skills predict agile practices maturity and not the other way around, we have the skills as independent variables and the agile practices measurements as dependent variables. If the ANOVA was significant at an alpha level of 5\%, we proceeded and built a multiple linear regression model to see which skills (factors) were significant. As a measurement of effect size we used $\eta^2$ (often called $R^2$ in regression analysis) for each omnibus test (i.e.\ ANOVA) \citep{coheneffect}.



\section{Results}\label{sec:results}
To assess the predictive power of non-technical individual skills on the agile practices, we ran eight independent ANOVA omnibus tests of which only the three response variables ``Iteration Planning'' ($F=2.166, p=0.017, N=105$), ``Customer Access'' ($F=2.415, p=0.008, N=102$), and ``Customer Acceptance Tests'' ($F=2.940, p=0.001, N=99$) were significant. All the other ANOVAs using ``Iterative Development'' ($F=1.307, p=0.224, N=103$), ``Continuous Integration and Testing'' ($F=0.664, p=0.792, N=100$), ``Stand-Up Meetings'' ($F=0.862, p=0.595, N=99$), ``Retrospectives'' ($F=0.946, p=0.510, N=101$), and ``Collocation'' ($F=0.968, p=0.488, N=103$) as dependent variables were not significant at an alpha level of 5\%. We therefore failed to reject the null-hypotheses in favour of H$_2$, H$_3$, H$_4$, H$_7$, H$_8$ using the agile practices as response variables, i.e.\ they were not significant at an alpha level of 5\%. Therefore, we also conclude that we have weak support, but still reject the null hypotheses in favour of H$_1$,H$_5$, and H$_6$. Next, we explain why the support was weak.

The agile practices Iteration Planning, Customer Access, and Customer Acceptance Tests had significant ANOVAs and we therefore ran further analyses using the significant factors (i.e.\ non-technical individual skills) in order to evaluate the size of the effects found. For these three significant omnibus tests, we built new models based on the significant factors and calculated effect sizes, which were found to be low or very low. The results were: Iteration Planning (adjusted $R^2=11.7\%$), using planning and teamwork skills as factors (see Table~\ref{planning}), Customer Access (adjusted $R^2=6.0\%$), using business-minded skills and organizing skills as factors (see Table~\ref{cust}), and finally Customer Acceptance Tests (adjusted $R^2=4.6\%$), using organizing skills as a factor (see Table~\ref{acc}). The only model with explained variance over ten percent was predicting ``Iteration Planning'' by using planning and teamwork skills skills. The regression models built using Customer Access and Customer Acceptance Tests had effect sizes under 10\%, showing low predictive power of non-technical individual skills, even though we rejected the null hypotheses.

Higher teamwork and planning skills were connected to better iteration planning, which is the only result that makes sense. Actually, organizing skills were negatively correlated to Customer Access and Customer Acceptance Tests, which questions the relevance of the results in general. This would then mean that good organizing skills would be bad for mature Customer Access and Customer Acceptance Tests practice, which seems odd. However, we believe these results also point out that individual satisfaction of e.g.\ organizing skills is a poor predictor of agile maturity of a practice. Additionally, these effect sizes should be seen as irrelevant according to \citet{coheneffect}, while above ten percent is considered only a small effect. Therefore, the significant connections between teamwork and planning skills and iteration planning are the only ones relevant to analyze further. 

The variance in the measurement of Iteration Planning, could be explained by 11.7\%, which is considered a small effect in these types of studies \citep{coheneffect}. As mentioned before, it makes sense that the individual skills of teamwork and planning would be connected to how well a team plans for an iteration, however, it makes equally much sense that individual communication, collaboration, decision-making, problem-solving, organizing, and negotiation skills would be connected to e.g.\ the team ability to develop iteratively. In addition, the effect was barely over ten per cent, which we still consider much lower that would be the case if all these agile team practices depended on individual non-technical skills.

\begin{table}
\renewcommand{\arraystretch}{1.5}
\caption{Linear Regression Coefficients (Dependent Variable: Iteration Planning with 112 valid cases). Adjusted $R^2=11.7\%$}
\label{planning}
\centering
\begin{tabular}{cccccccccccc}
\hline
\bfseries  Model & \bfseries Unstandardized B &\bfseries  Std. Error &\bfseries  Standardized B    &\bfseries  t  &\bfseries  $p$-value \\
\hline
(Constant) & 3.373 & 0.528  &  & 6.389 & 0.000* \\
\hline
Teamwork Skills & 0.175 &  0.074 & 0.778 & 2.354 & 0.020*   \\
\hline
Planning Skills & 0.133 &  0.062 & 0.208 & 2.148 & 0.034*   \\
\hline
{*p$<$.05}  \\
\end{tabular}
\end{table}

\begin{table}
\renewcommand{\arraystretch}{1.5}
\caption{Linear Regression Coefficients (Dependent Variable: Customer Access with 108 valid cases). Adjusted $R^2=6.0\%$}
\label{cust}
\centering
\begin{tabular}{cccccccccccc}
\hline
\bfseries  Model & \bfseries Unstandardized B &\bfseries  Std. Error &\bfseries  Standardized B    &\bfseries  t  &\bfseries  $p$-value \\
\hline
(Constant) & 5.251 & 0.692  &  & 7.590 & 0.000* \\
\hline
Business-Minded Skills & 0.173 &  0.080 & 0.204 & 2.150 & 0.034*   \\
\hline
Organizing Skills & -0.181 &  0.077 & -0.222 & -2.342 & 0.021*   \\
\hline
{*p$<$.05}  \\
\end{tabular}
\end{table}

\begin{table}
\renewcommand{\arraystretch}{1.5}
\caption{Linear Regression Coefficients (Dependent Variable: Customer Acceptance Tests with 104 valid cases). Adjusted $R^2=4.6\%$}
\label{acc}
\centering
\begin{tabular}{cccccccccccc}
\hline
\bfseries  Model & \bfseries Unstandardized B &\bfseries  Std. Error &\bfseries  Standardized B    &\bfseries  t  &\bfseries  $p$-value \\
\hline
(Constant) & 5.811 & 0.757  &  & 7.679 & 0.000* \\
\hline
Organizing Skills & -0.269 &  0.111 & -0.234 & -2.431 & 0.017*   \\
\hline
{*p$<$.05}  \\
\end{tabular}
\end{table}

\paragraph{Validating individual skills against perceived code quality}
As a validation of the non-technical individual skills measurements we also built a model using the perceived code quality in products as a response variable. The non-technical individual skills showed no connection to the agile team members' perceived code quality ($F=1.172, p=0.314, N=99$), which means that non-technical individual skills also failed to predict the agile team members' perceived quality of the code.

\paragraph{Summary of statistical results} To summarize the results above, we first analyzed our survey data for normality and plotted frequency histograms for multiple regression models using non-technical individual skills as factors and the agile practices (one-by-one), and perceived code quality, as response variables. To assess the predictive power of non-technical individual skills on the agile practice and perceived quality, we ran nine independent ANOVA omnibus tests of which only the three response variables ``Iteration Planning,'' ``Customer Access,'' and ``Customer Acceptance Tests'' were significant. For these three significant omnibus tests, we build new models based on the significant factors and calculate effect sizes, which were found to be low or very low. The only model with explained variance over ten percent was predicting ``Iteration Planning'' by using teamwork and planning skills. This result makes sense, however, so would many other predictions which were not significant, and in addition, higher organizing skills were connected to lower Customer Access and Customer Acceptance Tests, which questions the relevance of such a measurement. The data analyses therefore have shown that looking at non-technical individual skills is not the optimal level of analysis when wanting to predict agile maturity. What this means and the implications of the low predictive power of non-technical individual skills in connection agile practices will be discussed next.  

It is important to, again, highlight the differences between correlation\slash predictive and causal models and in this study we only claim the former.

\section{Discussion}\label{sec:discussion}
The results show that there is very little predictive power when using self-assessed non-technical individual skills to understand the perceived maturity of the agile practices. Based on previous work in software engineering we would \emph{not} expect such a result. This means that we can not look at self-assessed non-technical individual skills when trying to predict the intended and mature use of agile practices. 

If there is little value in looking at individual non-technical skills when understanding or improving agile practices, what is then the option? As mentioned in the introduction, \citet{turley1994identifying} identified 38 essential competencies of software engineers on different abstraction levels, which we interpret as an indication of the issues shown in our present study with using non-technical individual skills in order to increase the use of agile practices in software development teams, i.e.\ they are simply too many. In accordance with \citet{Tanner2014} and \citet{Crowder2015} who investigated agile project success in relation to team orientation, shared leadership, backup behavior, feedback, team autonomy, team learning, coordination, communication, trust, collective culture, team familiarity, customer involvement and mandate, stakeholder involvement and buy-in, and team structure and team logistics, we have also found empirical support for looking at other levels of analysis than the individual, when wanting to optimize the benefits from an agile approach. Since project success is connected to agility \citep{serrador2015does} and \citet{so} suggest a measurement for the intended use of the agile practices, we assume that higher scores on the agile practices measurement do imply a higher probability of project success. \citet{Lalsing2012} also state that it is of utter importance to find the ``right people for the right team'' and not only the ``right people,'' which seems to be a key when building teams that can leverage agile practices in the way that they are intended.

\citet{hackman2003learning} underlines the importance of crossing levels in organizational research and in this present study, we have shown that the individual level does not explain much variance and we instead need to investigate the team as the level of analysis. The cross-section between what is team and organization could be hard to define and e.g. organizational routines could be defined as both \citep{pentland2005organizational}. However, the distinctiveness between teams in organizations are often possible to find, and we argue the team-level needs to be in focus instead of the individual level, but preferably also in relation to the organizational level.

As a comparison, we looked at how \citet{edum2000developing} present individual skills needed in the construction industries. They suggest skills not far from what is suggested in software development, which supports our claim that these non-technical individual skills are too general to be useful in predictions of the dynamics of the specific organizational case. \citet{edum2000developing} divide the needed abilities into primary and secondary knowledge and skill elements for developing project management competencies. Among the primary knowledge and skill elements, they report: (1) Planning and scheduling, (2) Construction management activities, (3) Basic technical knowledge in own field, (4) Productivity and cost control, (5) Leadership, (6) Delegation, (7) Negotiation, (8) Decision making, (9) Motivation and promotion, (10) Team working, (11) Time management, (12) Top management relations, (13) Establishing budgets, (14) Reporting systems, (15) Drafting contracts, (16) Communication skills Presentation, (17) General and business correspondence, (18) Report writing, (19) Chairing meetings, and (20) Understanding of organization.

We believe our findings imply that the teams need all the abilities that the non-technical individuals skills try to capture, but should be seen, and investigated, as a team capacity instead. This means that there is a difference between an individual having team working skills and the skills the team as a whole possesses. Our present study supports the findings presented in Section~\ref{org}, that team skills are key to implementing and using agile practices. The collective intelligence is a property of the team itself \citep{woolley2010evidence} and, therefore, also the agile practices, i.e.\ just like the collective intelligence is unrelated to individual intelligence \citep{woolley2010evidence}, individual non-technical skills seem to be unrelated to non-technical team skills. In addition, personalities can be consciously changed over time \citep{hudson2015volitional} and depend on our group membership \citep{reynolds2001role}. From the software development context, \citet{grenjss2} present an interesting quote in their study saying that the interviewees were surprised by how vocal some, previously very quiet, programmers get on some of their agile teams. Such a finding has extensive empirical support from social psychology and the studies of how context and its social expectations and interactions form reality \citep{snyder1984belief}.

\subsection{Threats to Validity}\label{sec:validity}
We reflect on the threats to validity using internal, external, construct, and conclusion validity following the guidelines by \citet{Wohlin2000Experimentation}.

\paragraph{Internal Validity}
We only used the self-assessed (i.e.\ perceived) non-technical individual skills. It is difficult to determine the correlation of these agile team members' self-assessed skills to their actual skills, or peer-assessed skills. However, it has been proven in psychology research that people overestimate their skills systematically (see e.g.\ \citet{alicke1995personal}). Hence, we assume our self-assessment to be a valid measurement when building associative models with relative associations between variables. A potential other threat is the operationalization of ``skills'' into asking about the agile team members' satisfaction of their own skill. As mentioned in the method, the reason for using the personal satisfaction of a certain non-technical skills was the fact that, in agile teams, such skills are perceived as utterly important and a satisfaction of a skill should then be more related to a peer-evaluation then if asked to only rate their individual skills. Yet, we recognize that the reported individual skills could differ from ``real'' or the actual skills perceived by peers. 

We did not include participants from companies not following agile practices for a comparison between other types of software development work practices. The rational behind this was our focus on companies using agile practices and the fact that we prioritized having a high number of participants than comparing the results to participants from organizations not following agile practices. We would also have had to specify and measure other work practices, which then became out of scope for this study. However, we would not be surprised if non-technical individual skills would turn out to be weakly connected to other types of high performance work practices, both in other software development methods, but also in other fields. 





\paragraph{External validity}
The sampling in this study represents a convenience sampling procedure, as the authors of this paper had direct or indirect research connections to people within the organizations that chose participants to the study. To mitigate this threat, we tried to diversify the sample of participants to include as a representative sampling of agile teams as possible. We involved participants from seven organizations from different continents, different sizes of organizations, as well as the public and private sectors. In addition, in order to correlate skills to levels of agile maturity, the organizations, and the participating agile teams, were all on different agile maturity levels, both within and between organizations. Four organizations were based in Brazil and three organizations were from Europe, The Netherlands, including two companies and one public sector IT department. We therefore believe that our sample is representative for agile team members since we sampled from IT organizations of different types and sizes. Again, we aimed at looking for correlations, but allowing the skills measurements to co-vary, which makes such a sample appropriate. We also opted not to specify that we only wanted software developers to answer the survey, since we knew that many team members conducting software development work, might have other titles. We instead explicitly asked all members to respond, but told them to skip questions they could not assess. Therefore, our sample reflects all the agile team members that were involved in software development.

\paragraph{Conclusion validity}
Due to our large sample size of 113 survey responses, we had a decent sample for building our linear regression models. We also made sure all the assumptions were fulfilled when building such statistical models. 





\paragraph{Construct validity}
To ensure that we do not just base our measurements on our assumptions, we conducted a literature review on individual skills in software development and agile projects, as well as on agile practices. From this review, we presented the most related work we found and step-by-step derived representative non-technical individual skills as well as agile practices. We only included a measurement if we found evidence for it in at least two publications. In addition, and more importantly, we failed to reject most of our hypothesis, which was not our initial intention, of course, but our data turned out to provide us with different results than we expected, i.e.\ we reported our results and discussed it without trying to fish for significant $p$ values in data.  

A common construct validity threat is hypothesis guessing. Since the participants filled out their surveys on paper and totally anonymous (i.e.\ not even stating their role or gender), we believe they were given a chance to answer honestly. We also did not inform participants about any of our hypotheses, but instead introduced our research topic of finding drivers behind agile practices in general over a large sample from more than one country.

\section{Conclusions and Future Work}\label{sec:conclusions}
This paper set out to investigate the assumption that non-technical individual skills are positively connected to the mature use of agile practices. Through building a set of multiple linear regression models using a total of 113 survey responses, we analyzed the predictive power in measuring individual skills in relation to agile practices. We found that there is very low power in using non-technical individual skills to predict the maturity of agile practices in software development teams. We, therefore, conclude that looking at non-technical individual skills is not the optimal level of analysis when trying to understand and predict the use of agile practices in the software development context.


Future studies should focus more on the team-level when understanding the use of agile practices and build upon such theories when understanding the dynamics of agile teams in addition to trying to validated the result of the present study using external, or peer-assessed, measurements of skills and agile practices. In future studies, instead of asking questions about the non-technical individual skills, we would suggest items regarding non-technical team-level skills. For example, given that members of the team have enough talent and experience for the kind of work that is conducted, items suggested by \citet{wageman2005team} might be useful, like for example, whether everyone in the team has the special skills that are needed for team work. Future studies should also include the macro level of analysis in order to investigate if aspect of team agility could be explained on the organizational level, which there are indications of \citep{roth}.

\section*{Acknowledgements}
We would like to thank all the participating companies as well as colleagues helping out in different phases of this research.

\bibliographystyle{model5-names}
\bibliography{skillsagileIST}
\newpage

\end{document}